\documentclass[aps,prd,twocolumn,showpacs, superscriptaddress, longbibliography,nofootinbib]{revtex4-1}  

\usepackage{dcolumn}   
\usepackage{amsmath}
\usepackage{mathtools}
\usepackage[T1]{fontenc}
\usepackage{amssymb}
\usepackage{amsfonts}
\usepackage{amsthm}
\usepackage{mathrsfs}
\usepackage{slashed}
\usepackage[all]{xy}
\usepackage{tensor}
\usepackage{stmaryrd}
\usepackage{bm}
\usepackage[usenames,dvipsnames]{xcolor}
\usepackage[colorlinks=true,
    citecolor=Blue]{hyperref}

	\definecolor{coralred}{rgb}{1.0, 0.25, 0.25}
	\definecolor{junglegreen}{rgb}{0.16, 0.67, 0.53}

\newcommand{\op}{\mathcal{O}}
\newcommand{\vep}{\varepsilon} 
\newcommand{\lie}{\pounds}
\newcommand{\ep}{\epsilon}
\newcommand{\Bif}{\mathcal{B}}

\newcommand{\Hajicek}{H\'a\'{\j}i\v{c}ek~}

\newcommand{\beq}{\begin{equation}}
\newcommand{\eeq}{\end{equation}}

\begin{document}

\title{Virasoro hair and entropy for axisymmetric Killing horizons}

\author{Lin-Qing Chen}
\email{linqing.nehc@gmail.com}
\affiliation{Dean's Research Group, Okinawa Institute of Science and Technology, 1919-1 Tancha, Onna-son, Okinawa 904-0495, Japan}
\affiliation{Perimeter Institute for Theoretical Physics, 31 Caroline St. N, Waterloo, ON N2L 2Y5, Canada}
\affiliation{Centre for Quantum Information and Communication, \'Ecole polytechnique
de Bruxelles, CP 165, Universit\'e Libre de Bruxelles,1050 Brussels, Belgium}

\author{Wan Zhen Chua}
\email{wchua@perimeterinstitute.ca}
\affiliation{Perimeter Institute for Theoretical Physics, 31 Caroline St. N, Waterloo, ON N2L 2Y5, Canada}
\affiliation{Department of Physics and Astronomy, University of Waterloo, Waterloo, ON N2L 3G1, Canada}

\author{Shuwei Liu}
\email{sliu@perimeterinstitute.ca}
\affiliation{Perimeter Institute for Theoretical Physics, 31 Caroline St. N, Waterloo, ON N2L 2Y5, Canada}
\affiliation{Department of Physics and Astronomy, University of Waterloo, Waterloo, ON N2L 3G1, Canada}

\author{Antony J. Speranza}
\email{asperanz@gmail.com}
\affiliation{Perimeter Institute for Theoretical Physics, 31 Caroline St. N, Waterloo, ON N2L 2Y5, Canada}

\author{Bruno de S. L. Torres}
\email{bdesouzaleaotorres@perimeterinstitute.ca}
\affiliation{Perimeter Institute for Theoretical Physics, 31 Caroline St. N, Waterloo, ON N2L 2Y5, Canada}
\affiliation{Department of Physics and Astronomy, University of Waterloo, Waterloo, ON N2L 3G1, Canada}
\affiliation{Instituto de F\'{i}sica Te\'{o}rica, Universidade Estadual Paulista, S\~{a}o Paulo, S\~{a}o Paulo, 01140-070, Brazil}

\begin{abstract}
We show that the gravitational phase space
for the near-horizon region of a bifurcate, axisymmetric
Killing horizon in any dimension admits 
a  2D conformal symmetry
algebra with central charges proportional to the area. 
This extends the construction of [Haco et. al., JHEP 12, 098 (2018)] 
 to generic Killing horizons appearing in solutions
of Einstein's equations, and motivates a holographic
description in terms of a 2D conformal field theory.  
The Cardy entropy in such a field theory  agrees with the Bekenstein-Hawking entropy of the horizon, suggesting
a microscopic interpretation.  
\end{abstract}
\maketitle

\section{Introduction}

The Bekenstein-Hawking black hole entropy 
$S_\text{BH} = A/4G$ \cite{Bekenstein1972, Bekenstein1973a, Hawking1974}
presents a challenge to 
quantum gravity to provide a microscopic explanation.  
One proposal is that the entropy counts edge degrees of freedom
living on the horizon, and is controlled by boundary
symmetries \cite{Carlip1995, Carlip1995a, Teitelboim1996}.  This 
idea was strikingly realized in Strominger's derivation
of the BTZ black hole entropy, using the Cardy
formula for a conformal field theory (CFT) 
with the Brown-Henneaux central 
charge \cite{Strominger1998, Banados1992, Cardy1986, Brown1986c}.
Much subsequent work has been devoted to generalizing this 
construction to other contexts.

Carlip in particular demonstrated that the conformal symmetries were not
special to $\text{AdS}_3$ black holes; rather, they arise for 
generic Killing horizons. 
In all cases, postulating a CFT description led to 
agreement between the Cardy entropy and $S_\text{BH}$
\cite{Carlip1999c, Carlip1999b, Solodukhin1999, Park1999, Park2002, Koga2001, Silva2002a}.
Although very insightful,
certain aspects of Carlip's construction raised additional 
questions.  The symmetry generators had to satisfy periodicity
conditions whose justifications were obscure, and only 
a single copy of the Virasoro algebra was found, whereas 
the 2D conformal algebra consists of two copies,
$\text{Vir}_R\times\text{Vir}_L$ 
\cite{Carlip2020, Averin2020}. 
The Kerr/CFT correspondence \cite{Guica2009, Compere2012}
provided some clarity, 
by making the connection between near horizon symmetries and holographic
duality more explicit, allowing intuition from AdS/CFT to 
be applied.  As a byproduct, it also motivated a different choice 
of near-horizon symmetry generators whose periodicities followed from
the rotational symmetry of the Kerr black hole, thereby resolving one
issue in Carlip's original construction \cite{Carlip2011a,
Carlip2011b}.
Another significant advance came from
Haco, Hawking, Perry, and Strominger (HHPS) \cite{Haco2018},
who exhibited a full set of $\text{Vir}_R\times \text{Vir}_L$ 
symmetries for Kerr
black holes of  arbitrary nonzero spin.  This work was generalized
to Schwarzschild black holes using a different collection of symmetry
generators in \cite{Averin2020}.

The present work will demonstrate that arbitrary bifurcate, axisymmetric
Killing
horizons  possess a full set of conformal symmetries,
which act on edge degrees of freedom, 
or ``hairs'' \cite{Hawking2016, Hawking2017}, on the 
horizon. 
We emphasize how these symmetries arise from generic properties of the 
near-horizon geometry, clarifying the geometric origin of the symmetries and 
greatly extending the regime of applicability of similar soft hair constructions.  
Furthermore, we show that conformal
coordinates can always be found which  foliate the near-horizon
region by locally $\text{AdS}_3$ geometries,
giving rise to symmetry
vector fields satisfying a
$\text{Witt}_R\times \text{Witt}_L$ algebra 
in the vicinity of 
the horizon. 
The coordinates depend on two free parameters, $\alpha$ and  $\beta$,
which are related to the CFT temperatures $T_R$, $T_L$ using
properties of the near horizon vacuum.  

When the symmetries generated by the vector fields are implemented
canonically on the gravitational phase space, 
the algebra is extended to $\text{Vir}_R\times \text{Vir}_L$, 
with central
charges $c_R$ and $c_L$ determined in terms of the area and angular
momentum of the horizon according to (\ref{eqn:cR}) and (\ref{eqn:cL}).
Imposing a constraint on  $\alpha$ and $\beta$,  motivated by integrability
of the charges generating the symmetry, sets
the central charges equal to each other and 
proportional to the horizon area according to (\ref{eqn:cequal}).  We note that this 
choice of temperatures is a novel discovery of the present work, and differs from the choice 
made in \cite{Castro2010, Haco2018} for Kerr.  
More generally, (\ref{eqn:cequal}) applies for any choice of $\alpha$ and
$\beta$ when appropriate Wald-Zoupas terms are used to define
the quasilocal charges \cite{Wald2000c, Chandrasekaran2020}.  
The Cardy formula \cite{Cardy1986}
with the central charges (\ref{eqn:cequal}) reproduces
the entropy of the horizon, suggesting
a dual description in terms of a CFT.  This result
therefore motivates investigations into 
holography for arbitrary Killing horizons, including
the de Sitter cosmological horizon, and nonrotating 
and higher dimensional black holes.

\section{Near-horizon expansion}\label{NearHorizonSection} 
We are interested in the 
form of the metric near a bifurcate, axisymmetric
Killing 
horizon in a solution to Einstein's equations in dimension
$d\geq 3$.  
Axisymmetry means that, in addition to the horizon-generating
Killing vector $\chi^a$, there is a commuting, rotational
Killing vector $\psi^a$ with closed orbits.  
Axisymmetric horizons are of interest since, by the rigidity
theorems, all black hole solutions are of this form
\cite{Hawking1972a, Hollands2007, Emparan2008, Hollands2012}.
In situations where the horizon possesses more than one
rotational Killing vector, we simply single out one and proceed with
the construction.  

The  conformal symmetries of the horizon are found by first
constructing a system of  ``conformal coordinates'', 
designed to exhibit a locally
$\text{AdS}_3$
factor in the metric when  expanded near the 
bifurcation surface.  The asymptotic symmetries of this $\text{AdS}_3$
factor comprise the conformal symmetries of the horizon.
We first define Rindler coordinates near the bifurcation surface
using 
a construction of Carlip \cite{Carlip1999b}, suitably modified
to incorporate the additional rotational symmetry.  
The gradient of $\chi^2$ defines a radial vector
\begin{equation} \label{eqn:rho}
\rho^a = -\frac{1}{2\kappa} \nabla^a\chi^2,
\end{equation}
where $\kappa$ is the surface gravity of $\chi^a$.  On 
the horizon, $\rho^a$ and $\chi^a$ coincide, but off the horizon,
$\rho^a$ is independent.  The vectors $(\chi^a, \psi^a, \rho^a)$ mutually commute, 
and hence can form part of a coordinate
basis with coordinates $(t,\phi,r_*)$.  The remaining 
transverse
coordinates are denoted $\theta^A$.   
It is convenient to reparameterize the radial coordinate
\begin{equation} \label{eqn:x}
x = \frac1\kappa e^{\kappa r_*},
\end{equation}
which has the interpretation of proper geodesic distance to the bifurcation
surface to leading order near the horizon.  

In these coordinates, the near-horizon metric takes on Rindler
form, 
\begin{align}
ds^2 =&\; 
-\kappa^2 x^2 dt^2 +  dx^2 + \psi^2 d\phi^2 
+ q_{AB} d\theta^A d\theta^B  \nonumber\\
&\;- 2 x^2 dt \big(\kappa N_\phi d\phi  
+ \kappa N_A d\theta^A \big) + \ldots \label{eqn:Rind}
\end{align}
where the dots represent terms at $\mathcal{O}(x^2)$ or higher that  
do not enter the remainder
of the calculation (see appendix \ref{app:metric} for additional 
details on this expansion).  
Except for $\kappa$, all coefficients appearing in the 
above expansion are functions of $\theta^A$.


The conformal coordinates are now defined, 
in analogy to similar 
constructions in \cite{Haco2018, Castro2010}, 
as\footnote{In the case of Kerr, these coordinates 
are rescaled by a function of $\theta$ from the coordinates
used by HHPS; however, doing so does not substantially 
change the construction.  
Also, $\phi$ is the comoving 
angular variable, as opposed to the standard Boyer-Lindquist $\phi$.
See appendix \ref{app:Kerr} for details on the expansion for Kerr.  
}
\begin{align}
        w^+ &= xe^{\alpha \phi + \kappa t} \label{eqn:wp}\\
        w^- &= x e^{\beta\phi -\kappa t} \\
        y &= e^{\frac{\alpha + \beta}{2} \phi} \label{eqn:y}
\end{align}
(see appendix \ref{app:dS} for a concrete
realization in the example of de Sitter space, and appendix \ref{app:Kerr} for a discussion
of the Kerr).

\begin{figure}[t]
	\centering
  	\includegraphics[height=7.5 cm]{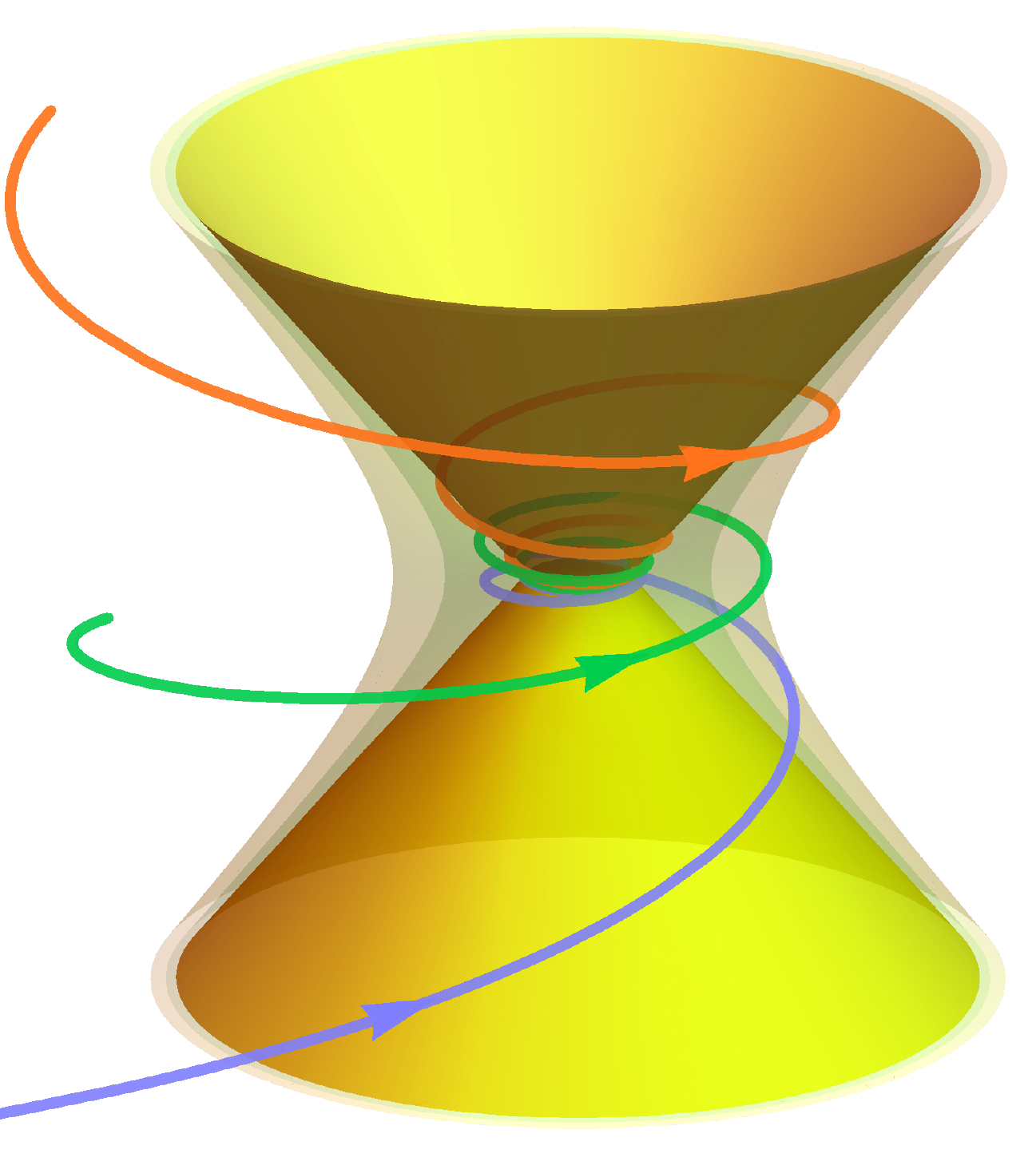}
	\caption{Plot of the bifurcate Killing horizon (yellow) with
	 three spirals depicting intersections of constant $w^+, w^-$. 
	 From top to bottom, they correspond to  $w^+/w^- = (5, 1, 1/5)$,  with the same value of the product $w^+ w^-$. They penetrate into any fixed-$x$ hyperbola,
	shown in light brown, at the same $y$. The arrows indicate the direction of increasing $y$ on approaching the bifurcation surface.}
	\label{Fig:wpfoliatef}
\end{figure}

Here, $\alpha$ and $\beta$ are free parameters that 
will later be related to 
the left and right temperatures of the 
system.
Because $x e^{\kappa t}$ and $x e^{-\kappa t}$ are simply the Kruskal
coordinates $V, U$ near the bifurcation surface, the 
future horizon is at $w^-=0$ and the past horizon is $w^+=0$ (see figure \ref{Fig:wpfoliatef} for a visualization of the conformal coordinates in the near horizon region).
Due to the periodicity $\phi \sim \phi + 2\pi$, 
the conformal
coordinates must be identified according to
\begin{equation} \label{eqn:period}
\left(w^+, w^-, y \right) \sim \left(e^{2\pi \alpha} w^+, 
e^{2\pi \beta}w^-, 
e^{\pi(\alpha + \beta)} y \right).
\end{equation}
  The near-horizon expansion in these coordinates becomes
\begin{align}
	ds^2   &=  \frac{dw^+ dw^-}{y^2} 
	+ \frac{4 \psi^2 }{(\alpha+\beta)^2 }\frac{dy^2}{y^2}
	+q_{AB}d\theta^A d\theta^B  \nonumber \\
	&\;	-\frac{2 dy}{(\alpha+\beta)y^3} \Big((\beta+ N_\phi  )   w^-dw^+ + (\alpha- N_\phi  )   w^+dw^-   \Big)  \nonumber \\
	&\; - \left(\frac{w^- dw^+}{y^2} - \frac{w^+ dw^-}{y^2}\right) \kappa N_A d\theta^A 
	+ \ldots \label{eqn:wpmmetric}
\end{align}
up to higher order terms in $w^+, w^-$. 
The first line takes the form of a locally $\text{AdS}_3$ metric
with a $\theta^A$-dependent
radius of curvature $\ell = \frac{2|\psi|}{\alpha+\beta}$, times a transverse metric.  
In intuitive words, the conformal coordinates zoom in on the near horizon region through the lens of  $e^{\alpha \phi}, e^{\beta\phi}$ attached to the Kruskal coordinate,  and bring out the  $\text{AdS}_3$ folia explicitly.

These coordinates allow for a
straightforward determination of the near horizon
symmetry generators.  They are the asymptotic
symmetry vectors of the $\text{AdS}_3$ factor in (\ref{eqn:wpmmetric}), 
where asymptotic refers to $y\rightarrow 0$.  
The vectors
are defined as in HHPS \cite{Haco2018} 
\begin{align}
\zeta^a_\vep = \vep(w^+)\partial_+^a +\frac12 \vep'(w^+) y\partial_y^a
\label{eqn:zep}\\
\xi^a_{\bar\vep} = \bar\vep(w^-) \partial_-^a +\frac12 \bar\vep'(w^-) y\partial_y^a\label{eqn:xep},
\end{align}
and one can readily verify that the Lie derivative of the first 
line of (\ref{eqn:wpmmetric}) with respect to these vectors vanishes 
up to $\op(y^{-3})$ terms. 
{\it A priori}, $\vep(w^+)$ and $\bar \vep(w^-)$ are arbitrary functions,
but in light of the
periodicity condition (\ref{eqn:period}),  the vector fields are 
single-valued only when
$\vep(w^+ e^{2\pi\alpha}) = \vep(w^+) e^{2\pi \alpha}$, 
$\bar \vep(w^-e^{2\pi\beta}) = \bar \vep(w^-) e^{2\pi\beta}$. 
A basis for such functions is 
\begin{align}
\vep_n(w^+) &= \alpha\, (w^+)^{1+\frac{in}{\alpha}},  \\
\bar \vep_n(w^-) &= -\beta\, (w^-)^{1-\frac{in}{\beta}},
\end{align}
and their corresponding generators will be labeled as $\zeta_n^a$, 
$ \xi_n^a$.  The algebra satisfied by these vector fields upon 
taking Lie brackets is two commuting copies 
of the Witt algebra, 
\begin{align}
[\zeta_m , \zeta_n] &= i(n-m)\zeta_{m+n} \\
[\xi_m ,\xi_n] &= i(n-m)\xi_{m+n}.
\label{eqn:zetxi}
\end{align}
The generators are defined in a neighborhood of the bifurcation
surface, but   
oscillate wildly as it is approached.  The $\zeta^a_n$ generators are regular on the future horizon but not the past, and similarly 
$\xi^a_n$ are regular at the past horizon, but not the future.  
The zero mode generators $\zeta_0^a$ and $\xi_0^a$
are regular everywhere, given by two helical
Killing vectors
\begin{align} \label{eqn:zet0}
\zeta_0^a &= \frac{\alpha}{\alpha+\beta}
\left(\frac{\beta}{\kappa}\chi^a + \psi^a\right) \\
\xi_0^a &= \frac{\beta}{\alpha+\beta} \left(\frac{\alpha}{\kappa}
\chi^a
-\psi^a\right).
\label{eqn:xi0}
\end{align}

The expressions (\ref{eqn:zet0}) 
lead to the interpretation of $\alpha$ and $\beta$
in terms of the right and left temperatures.  The analog
of the Frolov-Thorne vacuum \cite{Frolov1989} for quantum fields 
near the bifurcation surface
is thermal with respect to the $\chi^a$ Killing vector.
The density matrix is therefore of the form
 $\rho\sim \exp\left(-\frac{2\pi}{\kappa} \omega_\chi\right)$, where 
$\omega_\chi = -k_a\chi^a$ is the frequency with respect to $\chi^a$ 
for a wavevector $k_a$. 
Reexpressing it in terms of $\zeta_0^a$, $\xi_0^a$ frequencies
via
\begin{equation}
\omega_\chi = 
-k_a\left(\frac{\kappa}{\alpha} \zeta_0^a 
+ \frac{\kappa}{\beta} \xi_0^a\right)
= \frac{\kappa}{\alpha}\omega_R+\frac{\kappa}{\beta}\omega_L
\end{equation}
shows that $\rho \sim \exp\left(-\frac{2\pi}{\alpha}\omega_R
-\frac{2\pi}{\beta}\omega_L\right)$, allowing us to read
off the temperatures $(T_R, T_L) 
= (\frac{\alpha}{2\pi},\frac{\beta}{2\pi})$ as the 
thermodynamic potentials conjugate to $\zeta_0^a, \xi^a_0$ \cite{Guica2009}.

\section{Central charges}

Having identified the near-horizon symmetry generators
(\ref{eqn:zep}) and (\ref{eqn:xep}), the next step is to 
implement them on the gravitational phase space.  
This involves identifying Hamiltonians $H_n, \bar H_n$
that generate the symmetries associated with $\zeta_n^a$, $\xi_n^a$,
 meaning
 \beq
\delta H_n = \Omega\big(\delta g_{ab}, \lie_{\zeta_n} g_{ab}\big)
\label{eqn:dHn}
\eeq
where $\Omega$ is the symplectic form of the phase space. 
 
Assuming integrable Hamiltonians can be found, their Poisson brackets
 automatically reproduce the algebra satisfied by the vector fields (\ref{eqn:zetxi}), up to central extensions,
\begin{align}
\{H_m, H_n\} &= -i\Big[(n-m) H_{m+n} + K_R(m,n)\Big] \label{eqn:HH}\\
\{\bar H_m, \bar H_n\}  & = -i\Big[(n-m)\bar H_{m+n} 
+ K_L(m,n)\Big]\label{eqn:bHbH} 
\end{align}
Since the Witt algebra has a unique nontrivial central extension to Virasoro,
 the central terms in the above expression must 
be of the form
\begin{align}
K_{R,L}(m,n) = \frac{c_{R,L}}{12} (m^3-m) \delta_{m+n,0} \label{eqn:KRL}
\end{align}
where the constants $c_{R,L}$ are the central charges.

Using the covariant phase space formalism 
and standard Iyer-Wald identities
\cite{Crnkovic1987, Lee1990a, Iyer1994a, Harlow2019}, 
the right hand side of 
(\ref{eqn:dHn}) can be expressed on-shell as 
\beq\label{eqn:IWident}
\Omega\big(\delta g_{ab}, \lie_{\zeta_n} g_{ab}\big)
= \int_{\partial \Sigma} (\delta Q_{\zeta_n} - i_{\zeta_n}\theta)
\eeq
where the integral is over 
the boundary of a Cauchy surface $\Sigma$ for the exterior region of the Killing
horizon.
The other quantities appearing in (\ref{eqn:IWident}) are the 
Noether potential $(d-2)$-form
\beq \label{eqn:IzetOm}
Q_{\zeta_n}
= -\frac{1}{16\pi G} \ep\indices{^a_b } \nabla\indices{_a} \zeta^b_n
\eeq
and the symplectic potential $(d-1)$-form,
\beq
\theta = \frac{1}{16\pi G} \ep^a\left(\nabla^b\delta g_{ab}-
g^{bc} \nabla_a \delta g_{bc} \right).
\eeq
In these expressions,
$\epsilon$ denotes the spacetime volume form, with uncontracted 
indices not displayed. 

The zero mode generators $\zeta_0^a, \xi_0^a$ are Killing 
vectors whose corresponding Hamiltonians are
\begin{align}
H_0 &= \frac{\alpha}{\alpha+\beta} \left(\!\frac{\beta A}{8\pi G}
+ J_H \right) \\
\bar H_0 &= \frac{\beta}{\alpha+\beta} 
\left(\! \frac{\alpha A}{8\pi G} - J_H \right),
\end{align}
where $A$ is the area of the bifurcation surface $\Bif$, and 
\begin{equation}\label{eqn:JH}
J_H\equiv \int_{\Bif} Q_\psi = \frac{1}{4G} \int d\theta^A \sqrt{q}|\psi|
N_\phi
\end{equation}
is the angular momentum of the Killing horizon.  $J_H$ 
agrees with the total angular momentum $J$ in an asymptotically 
flat vacuum solution, but generally differs when matter
is present outside the horizon \cite{Poisson2004}.  
The remaining generators with $n\neq 0$ vanish in the Killing horizon background,
because the vector fields $\zeta_n^a$, $\xi_{-n}^a$ have an $e^{in\phi}$ angular dependence,
which integrates to zero on the axially symmetric horizon.  Of course, the variations of these 
other generators are nonzero.

The Poisson bracket only involves variations of the Hamiltonians, and hence
can be computed directly from (\ref{eqn:dHn}).
According to (\ref{eqn:HH}) and (\ref{eqn:KRL}),
the central charge  appears as the coefficient of the $m^3$ term in $\{H_m, H_{-m}\}$.  
Because of the singular limit in the generators $\zeta_m^a$,  
the integral that computes the bracket cannot be evaluated directly on the bifurcation surface.  
Instead, we work on  a cutoff surface at constant $x$ and $t$, and 
perform the integration before taking the limit $x\rightarrow0$.  
The details of this calculation are given in appendix \ref{app:cpp}, and 
results in the central charge 
\begin{equation} \label{eqn:cR} 
c_R = \frac{24}{(\alpha+\beta)^2} \left(\frac{\beta A}{8\pi G}
+ J_H\right) .
\end{equation}
An analogous calculation for $\{\bar H_m,\bar H_{-m}\}$
yields 
\begin{align}
c_L &= \frac{24}{(\alpha+\beta)^2} \left(\frac{\alpha A}{8\pi G}
-J_H\right) . \label{eqn:cL} 
\end{align}

\section{Temperature and Entropy} 

Up to now, we have carried out the full calculation with  arbitrary temperatures. 
However, experience with asymptotic symmetries in $\text{AdS}_3$ \cite{Brown1986c}
suggests that the left and right central charges should be equal.  By equating
(\ref{eqn:cR}) and (\ref{eqn:cL}), we arrive at the condition 
\beq\label{tep}
\alpha - \beta = \frac{16\pi G J_H}{A},
\eeq
which, notably, differs from the choice of temperatures employed by HHPS for Kerr
black holes \cite{Castro2010, Haco2018}.  
With this constraint, the central charges are proportional to the horizon
area,
\begin{equation} \label{eqn:cequal}
c_R=c_L = \frac{3 A }{2\pi G(\alpha+\beta)  }  .
\end{equation}  
It is natural to conjecture that the condition (\ref{tep})
arises from imposing appropriate boundary conditions that ensure the charges $H_n$, $\bar H_n$
are integrable.  

This conjecture turns out to be correct, as was recently demonstrated in 
\cite{Chandrasekaran2020}, which carried out a systematic analysis of $\theta$ 
appearing in (\ref{eqn:IWident}) pulled back to the future or past horizon.  
The main point is that the part of $\theta$ that can be written as a total
variation, $-\delta \ell$,
contributes to the Hamiltonians $H_n, \bar H_n$, and in order for 
the values computed for 
$H_0, \bar H_0$ on the past horizon to agree with their values computed on the future horizon, 
the expression
for $\ell$ must be the same on each horizon.  Furthermore, reference \cite{Chandrasekaran2020}
showed that 
the central charges can 
be expressed  in terms of $\ell$ and its variation, and since the same quantity is used on the past 
and future horizon, one can derive that the central charges must be equal.
On the other hand, when the charges are integrable, the calculations leading to 
(\ref{eqn:cR}) and (\ref{eqn:cL}) remain valid, and hence $\alpha$ and $\beta$ must be
chosen to ensure that $c_R=c_L$, which produces equation (\ref{tep}).

The results of \cite{Chandrasekaran2020} further allow us to work out the boundary conditions 
that need to be imposed at the future $\mathcal{H}^+$
or past horizon $\mathcal{H}^-$ to arrive at integrable generators.\footnote{For a derivation
that explores alternative boundary conditions, 
see \cite{Chen2020}.}
One of the conditions is local,
\beq \label{eqn:cond1}
g^{ab}\delta g_{ab} \overset{\mathcal{H}^\pm}{=}0,
\eeq
while the second condition is a weaker, integrated condition over the transverse
$\theta^A$ directions,
\beq\label{eqn:cond2}
\int d\theta^A \sqrt{q}|\psi|\left(\delta k -k q^{ab}\delta g_{ab} +2 \varpi^a\chi^b\delta g_{ab} \right)
=0,
\eeq
where $k$ is the inaffinity,\footnote{Although in the background 
$k$ is the same as the surface gravity $\kappa$, we use a different letter to allow $\kappa$ to be 
fixed constant, while $\delta k\neq 0$.} 
defined by $\chi^a\nabla_a \chi^b = k \chi^b$,
$\varpi_b = -q\indices{^a_b}n^c\nabla_a \chi_c $ is the \Hajicek one-form, $n^c$ is an auxiliary 
transverse null vector to the horizon, and $q^{ab} = g^{ab} +n^a \chi^b +\chi^a n^b$.
The variations in this expression are taken holding $\chi_a$ fixed; see appendix \ref{app:WZ}
for additional details on these quantities and their variations.

Since the near-horizon gravitational phase space exhibits 
$\text{Vir}_R\times \text{Vir}_L$ symmetry, 
considerations from holographic duality suggest 
that its quantum  description
is given by a 2D CFT.
In such a theory, unitarity and modular invariance determines the 
asymptotic density of states via the Cardy formula.
Using the temperatures $(T_R, T_L) = (\frac{\alpha}{2\pi}, \frac{\beta}{2\pi})$ derived from properties of the Frolov-Thorne
vacuum, and central charges (\ref{eqn:cequal}), the 
Cardy formula for the canonical ensemble yields an entropy
\cite{Cardy1986, Compere2019}
\begin{equation} \label{eqn:SCardy}
S = \frac{\pi^2}{3}(c_R T_R + c_L T_L) = \frac{A}{4 G},
\end{equation}
which agrees with the Bekenstein-Hawking entropy.  This 
therefore motivates
an interpretation of the black hole
microstates in terms of a dual CFT.

\section{Wald-Zoupas term}

Instead of imposing the boundary conditions
(\ref{eqn:cond1}) and (\ref{eqn:cond2}), we could instead work with nonintegrable 
charges that are not conserved due to a loss of symplectic flux from the region outside the 
horizon.  In this case, the Wald-Zoupas procedure gives a suitable definition of the 
quasilocal charges \cite{Wald2000c}.  This prescription corrects $\delta H_n$ by a flux contribution,
constructed from terms appearing in (\ref{eqn:cond1}) and (\ref{eqn:cond2}) that the boundary 
condition would have set to zero.  
The bracket of the charges must then be modified, 
in which case the 
the Barnich-Troessaert bracket provides a suitable 
definition \cite{Barnich2011c}.  Doing so shifts the central charges by
\begin{equation} \label{eqn:DcR}
\Delta c_R =
\frac{-12}{(\alpha+\beta)^2} \left((\beta-\alpha)\frac{A}{8\pi G} + 2 J_H\right)
\end{equation}
and $\Delta c_L = -\Delta c_R$ (see appendix \ref{app:WZ} for details).  
Adding these to (\ref{eqn:cR}) and (\ref{eqn:cL}) sets the two central
charges equal, and given by (\ref{eqn:cequal}), but now with any
choice of $\alpha$ and $\beta$.  Hence, the choice of $\alpha-\beta$
described in (\ref{tep}) is also the unique choice which sets the 
Wald-Zoupas corrections to the central charges to zero.  
Additional details of this Wald-Zoupas prescription are given in \cite{Chandrasekaran2020}.

The central charges (\ref{eqn:cequal}) can be compared to those 
found by HHPS \cite{Haco2018},
whose choice of temperatures for the Kerr 
black hole set $\alpha+\beta = \frac{A}{8\pi G J_H}$.  
Substituting this into (\ref{eqn:cequal}) reproduces 
their result $c_R = c_L = 12 J_H$.  Our results are therefore consistent
with theirs, although we have 
demonstrated that once the Wald-Zoupas terms are included, the 
construction does not appear to rely on any specific choice of 
temperatures.

\section{Discussion}

The agreement between the  horizon entropy and the Cardy formula with 
central charges (\ref{eqn:cequal}) suggests 
that the quantum description of the horizon involves
a CFT.  A conservative interpretation of
this result is that the presence of the horizon breaks
some gauge symmetry of the theory, giving rise to edge degrees of 
freedom \cite{Carlip1995, Carlip1995a, Teitelboim1996}.  
The $\text{Vir}_R\times \text{Vir}_L$ algebra then provides a symmetry principle that 
constrains the quantization of these edge modes, which is strong enough to determine the 
asymptotic density of states accounting for the entropy.  This argument holds even if 
the conformal symmetries are only a subset of the full horizon symmetry algebra. 
Other horizon symmetries 
can include additional rotational symmetries, supertranslations, and diffeomorphisms  of the 
bifurcation surface
\cite{Hawking2016, Hawking2017, Donnay2016, Donnay2016a, Donnelly2016a, Speranza2018a, Carlip2018, Carlip2020, Chandrasekaran2018, 
Ciambelli2019a, Donnay2019, Compere2019, Grumiller2020, Adami2020}, and determining how they interact with the conformal
symmetries of this paper would be an interesting direction to pursue.  It is also possible that a slightly different 
symmetry algebra can be used to fix the entropy; in particular, \cite{Aggarwal2020} showed that the HHPS 
construction can be modified to produce a Virasoro-Kac-Moody
symmetry characteristic of a warped CFT.  A straightforward alteration of our construction should demonstrate how to realize warped conformal symmetries on arbitrary
axisymmetric Killing horizons.  

A more ambitious proposal is that the near-horizon region is 
holographically dual to a CFT. This is in line with the Kerr/CFT correspondence \cite{Guica2009, Compere2012},
and raises the exciting possibility of producing new interesting examples of holography
for a variety of different Killing horizons.  In this picture, the expression (\ref{eqn:cequal})
can be interpreted as determining the horizon area in terms of the 
temperatures $(\frac{\alpha}{2\pi}, \frac{\beta}{2\pi})$ and the central charge.  
A rather nontrivial aspect of such a proposed duality, however, is the lack of a decoupling 
limit for the near-horizon region, due to nonextremality, $\kappa\neq 0$.  
The anticipated need for Wald-Zoupas terms in defining integrable charges can be viewed as 
one indicator of this lack of decoupling, since they imply a loss of 
symplectic flux from the subregion under consideration.  The CFT should therefore be  an open quantum system,  deformed 
by an operator coupling it to an auxiliary system describing the far away region.  
This is quite reminiscent of recent models of black hole evaporation in holography \cite{Almheiri2019, Penington2019,
Almheiri2020,Penington2019a}, 
and hence studying the holographic
description of Killing horizons may lead to new insights on the black hole information problem.

These results open a number of directions for further 
investigation. 
The parameters $\alpha$ and $\beta$ 
were not fully fixed by the arguments in this paper; even the 
condition (\ref{tep}) does not determine 
the sum $\alpha + \beta$.  With the Wald-Zoupas terms, any choice 
of $\alpha$ and $\beta$ leads to the correct Cardy entropy, and 
so it remains to be seen what other physical requirements fix 
their value.  It may be that any choice is valid, which has some advantages because it can be used to ensure $1\ll c_{R,L} \ll H_0,\bar{H}_0$, which is the regime in which the Cardy formula is valid.   The temperatures used by HHPS  were determined using 
the hidden conformal symmetry of scalar scattering amplitudes in the near region of
Kerr \cite{Castro2010}, and we are currently investigating its implication in our construction.

A natural question is whether this construction works for other types
of horizons or subregions.  
With mild modifications, we expect it to 
work for degenerate Killing horizons with $\kappa=0$.  
One could also consider noncompact horizons, such as Rindler
space, in which the vector field $\psi^a$ does not 
have closed orbits.  In these cases, one could quotient
by a finite translation along $\psi^a$, which serves to 
both regulate the horizon area and to impose periodicity
conditions on the generators.  This should lead to a sensible
notion of entropy density following from the Cardy formula.  
Other possible subregions to consider are cuts of a Killing 
horizon or more generic null surfaces
\cite{Chakraborty2016, Chandrasekaran2018, Ciambelli2019a, Grumiller2020, Adami2020}, conformal Killing horizons \cite{Jacobson2019}, causal diamonds
\cite{Jacobson2016}, 
Ryu-Takayanagi surfaces \cite{Ryu:2006bv, Ryu:2006ef}, 
and generic subregions \cite{Donnelly2016a, Speranza2018a}.

\acknowledgments
We acknowledge Niayesh Afshordi, Glenn Barnich, Venkatesa Chandrasekaran, Luca Ciambelli, 
St\'ephane Detournay, Ted Jacobson, Rob Leigh, Alex Maloney, Dominik Neuenfeld, Lee Smolin, Jie-Qiang Wu, Beni Yoshida, and C\'eline Zwikel for helpful
discussions, and William Donnelly for comments on a draft of this work.   
We thank  Malcolm Perry for suggesting the question of generalizing the HHPS construction to the cosmological horizon. LQC is very grateful to Vyacheslav Lysov for  insightful conversations on this theme for over a year. We thank Geoffrey Comp\`ere  and Adrien Fiorucci for providing the SurfaceCharges Mathematica package \cite{Compere2012}. 
We are very grateful to the organizers of the Perimeter Scholars Institute 
Winter School, where this work was initiated.
 AJS thanks the Kavli Institute for Theoretical Physics for hospitality during the Gravitational Holography program.
 LQC thanks Perimeter Institute for the hospitality during the 2019-2020 visit. Research at Perimeter Institute is supported in part by the Government of Canada through the Department of Innovation, Science and Economic Development Canada and by the Province of Ontario through the Ministry of Colleges and Universities. LQC thanks Okinawa Institute of Science and Technology for academic visiting funding. BSLT thanks IFT-UNESP/ICTP-SAIFR and CAPES for partial financial support. 
This research was supported in part by the National Science Foundation under Grand No. NSF PHY-1748958.


\appendix
\section{Near-horizon coordinates}
\label{app:metric}

This appendix provides additional details
leading to the Rindler expansion of the 
near-horizon metric (\ref{eqn:Rind}).
We set $\gamma^a = \partial_x^a$ to be 
the radial coordinate vector in the $x$ coordinate system; using (\ref{eqn:x})
it is related to $\rho^a$ by
\begin{equation} \label{eqn:gammaa}
\gamma^a= \frac{1}{\kappa x} \rho^a.
\end{equation}
To see that $x$ agrees with the proper
distance to the bifurcation surface at leading 
order, first note
that the norms of $\rho^a$ and $\chi^a$ are 
related by 
$\rho^2 = -\chi^2 +\op(\chi^4)$ \cite{Carlip1999b}, which
implies that $\partial_{r^*}(\chi^2) = \rho^a\nabla_a\chi^2
= 2\kappa \chi^2 +\op(\chi^4)$, and hence 
$\chi^2 = -e^{2\kappa r^*}$
to leading order near the horizon.  Then we find that 
\begin{equation} \label{eqn:chi2}
\chi^2 = -\kappa^2 x^2 + \op(x^4)
\end{equation}
and 
\begin{equation}
\gamma^2 = 1+\op(x^2)
\end{equation}
showing that $\gamma^a$ is unit normalized near the horizon, 
which gives its parameter $x$ the interpretation of the proper 
distance.  

Equations (\ref{eqn:gammaa}) and 
(\ref{eqn:chi2}) can be used to obtain more information
about the near horizon expansion of the metric.  Expressing the 
$\op(x^4)$ term as $-2 x^4  \kappa^2 M(\theta^A)$, 
where $\theta^A$ are transverse
coordinates on constant-$(t,x,\phi)$ surfaces, we find that 
\begin{equation}
\gamma_a = \frac{-1}{2\kappa^2 x} \nabla_a \chi^2 = 
\nabla_a x\left(1+4 x^2  M\right)
+x^3\nabla_a M +\op(x^4).
\end{equation}
This expression determines the expansion of the $g_{x\mu}$ components 
of the metric.  Note that by definition, $\chi^a\gamma_a = 
\psi^a\gamma_a = 0$, and so  $g_{tx}$ and $g_{\phi x}$  
identically vanish in this coordinate system.  
The coordinates $(t,x,\phi)$ are defined up to shifts
by functions of $\theta^A$, although demanding that 
$x=0$ coincides with the bifurcation surface eliminates
the shift freedom for $x$.  By shifting 
$\phi\rightarrow \phi + G(\theta^A)$, we can arrange for $g_{\phi A}$
to vanish on the bifurcation surface. Note that this may 
spoil manifest invariance with respect to other symmetries, for example, in a Myers-Perry black hole in with angular momentum in other directions besides the $\psi^a$ rotation \cite{Myers1986, Myers2011}. 
Furthermore, a bifurcate Killing horizon exhibits a discrete
reflection symmetry through the bifurcation surface, which
to leading order sends $x\rightarrow -x$ \cite{Kay1991}.  
This can be used to 
rule out any terms appearing at $\op(x)$ in the near horizon expansion
of the metric.  These considerations result in the following form
of the near horizon metric,
\begin{align}
ds^2 =&\; 
-\kappa^2 x^2 dt^2 + \big(1+4 x^2 M\big) dx^2 + \psi^2 d\phi^2 
+ q_{BC} d\theta^B d\theta^C  \nonumber\\
&\;- 2 x^2 dt \big(\kappa N_\phi d\phi  
+ \kappa N_B d\theta^B \big) + 2 x^3  \partial_B M\, d\theta^B dx \nonumber  \\
&\; + x^2\left(\psi_{(1)}^2 d\phi^2 + 2 \Psi_B d\phi d\theta^B + q^{(1)}_{BC}d\theta^B d\theta^C \right) \nonumber \\
&\;+ \op(x^4) 
\end{align}
where all the coefficients are functions of 
$\theta^A$, with the exception of $\kappa$, which is constant
by the zeroth law of black hole mechanics.  Note that $\psi^2$ 
coincides with the squared norm of the rotation Killing vector $\psi^a$ on the bifurcation surface.

The transformation between Rindler and 
conformal coordinates (\ref{eqn:wp}-\ref{eqn:y}) can be  
inverted to obtain
\begin{align}
        t &= \frac1{2\kappa} \log\left[\frac{w^+}{w^-} 
        y^{2\left(\frac{\beta-\alpha}{\beta+\alpha}\right) } \right]\\
        x &= \left(\frac{w^+ w^-}{y^2}\right)^{1/2}\\
        \phi &= \frac{2}{\alpha+\beta} \log y.
\end{align}
These inverse transformations are useful when
expressing the generators $\zeta_n^a, \xi_n^a$
in terms of $\chi^a, \psi^a$, and $\rho^a$ in
(\ref{eqn:znkilling}), (\ref{eqn:xinkilling}),
and are helpful when trying to understand
how the $\text{AdS}_3$ folia embed into the 
near-horizon region.  
Using this inverse transformation, it is straightforward to 
express the symmetry generators (\ref{eqn:zep}) and (\ref{eqn:xep}) 
in terms of $\rho^a$ 
and the Killing vectors $\chi^a$, $\psi^a$,
\begin{align}
\zeta_n^a&= 
(w^+)^{\frac{in}{\alpha}}
\left[\zeta_0^a - \frac{in\rho^a}{2\kappa}+ \frac{in}{\alpha+\beta}
\left(\frac{\beta-\alpha}{2\kappa}\chi^a+\psi^a\right)
\right]  \label{eqn:znkilling}\\
\xi_n^a&= 
(w^-)^{\frac{-in}{\beta}}
\left[\xi_0^a -\frac{in\rho^a}{2\kappa}+ \frac{in}{\alpha+\beta}
\left(\frac{\beta-\alpha}{2\kappa}\chi^a+\psi^a \right)
\right] \label{eqn:xinkilling}
\end{align}
with $\zeta_0^a, \xi_0^a$ given in (\ref{eqn:zet0}) and (\ref{eqn:xi0}).

\section{de Sitter example} \label{app:dS}

In the course of this work, the abstract construction described in \ref{NearHorizonSection} was largely inspired by first working out the case of $4D$ de Sitter space. It was later realized that a similar strategy could also be applied to Kerr, and only then was the generalization to any bifurcate axisymmetric Killing horizon found. In static coordinates $(t, r, \theta, \phi)$, the de Sitter metric reads

\begin{equation}\label{desittermetric}
    ds^2 = -\left(1 - \dfrac{r^2}{\ell^2}\right)dt^2 + \dfrac{dr^2}{\left(1 - r^2/\ell^2\right)} + r^2 d\Omega^2_2,
\end{equation}
where $d\Omega^2_2 = d\theta^2 + \sin^2\theta d\phi^2$. In these coordinates, one makes
explicit the presence of a \emph{cosmological horizon} at $r = \ell$, which is a Killing
horizon with generator \mbox{$\chi^a = (\partial_t)^a$}, and surface gravity $\kappa = \ell^{-1}$. The azimutal Killing vector is $\psi^a = (\partial_\phi)^a$, and the vector $\rho^a$ is 

\begin{equation}
    \rho^a = -\dfrac{r}{\ell}\left(1 - \dfrac{r^2}{\ell^2}\right)\partial_r^a = \partial_{r_\ast}^a,
\end{equation}
where we can define the tortoise coordinate $r_\ast$ at lowest order near $r = \ell$ to be

\begin{equation}
    r_\ast \simeq \ell \log\sqrt{2\left(1 - \dfrac{r}{\ell}\right)}.
\end{equation}
This then implies, from the definition in Equation \eqref{eqn:x},
\begin{equation}
    x \simeq \sqrt{2\ell\left(\ell - r\right)}
\end{equation}
and that is precisely the proper radial distance to the horizon, up to lowest nontrivial order.

Expressing the metric \eqref{desittermetric} with the tortoise coordinate $r_\ast$, we have near the horizon

\begin{equation}
    ds^2 \simeq \left(1 - \dfrac{r^2}{\ell^2}\right)\left(- dt^2 + dr^2_\ast\right) + r^2d\Omega^2_2,
\end{equation}
and therefore radial null geodesics will approach the horizon with $t - r_\ast = \text{const}$ (outgoing geodesics) or $t + r_\ast = \text{const}$ (incoming geodesics). We can then parametrize the approach to the horizon with finite coordinate values by defining $w^{\pm}$ as functions of the form $f(r_\ast \pm t)$, respectively; further imposing that $w^{\pm}$ is proportional to $x$ for constant $t$ fixes the function $f(r_\ast \pm t) \propto e^{(r_\ast \pm t)/\ell}$, and adding the periodicity condition via an exponential dependence on the azimuthal angle $\phi$ finally leaves us with

\begin{equation}
\begin{gathered}
    w^+ = \sqrt{2\ell\left(\ell - r\right)}e^{\alpha\phi + t/\ell},\\
    w^- = \sqrt{2\ell\left(\ell - r\right)}e^{\beta\phi - t/\ell},\\
    y = e^{\frac{\alpha + \beta}{2}\phi}.
\end{gathered}
\end{equation}

\section{Kerr example} \label{app:Kerr} 
The analogous construction illustrated in~\ref{app:dS} in the case of Kerr is also possible. In Boyer-Lindquist coordinates $(t, r, \theta, \Tilde{\phi})$, the Kerr metric is

\begin{align}
    ds^2 =& -\left(1 - \dfrac{2 G M r}{\Sigma}\right)dt^2  - \dfrac{4 G M r a \sin^2\theta}{\Sigma}dt d\Tilde{\phi} \nonumber \\
    & + \dfrac{\sin^2\theta}{\Sigma}\Big((r^2+a^2)^2 - a^2\Delta \sin^2\theta\Big)d\Tilde{\phi}^2 \nonumber \\
    & + \dfrac{\Sigma}{\Delta}dr^2 + \Sigma d\theta^2,
\end{align}
where
\begin{equation}
    \begin{gathered}
    \Delta = r^2 + a^2 - 2 G M r = (r-r_+)(r-r_-),\\
    \Sigma = r^2 + a^2\cos^2\theta.
    \end{gathered}
\end{equation}
 We denote the azimuthal angle of Boyer-Lindquist coordinates by $\Tilde{\phi}$ to emphasize its distinction from the co-rotating angle $\phi = \Tilde{\phi} - \Omega_H t$ used in the main body of the paper. The horizon-generating Killing vector and the corresponding surface gravity for the outer horizon at $r = r_+$ are
 
 \begin{align}
     \chi^a = (\partial_t)^a + \Omega_H (\partial_{\Tilde{\phi}})^a, \\
     \kappa = \dfrac{(r_+ - r_-)}{2(r_+^2 + a^2)},
 \end{align}
 with $\Omega_H = a/(r_+^2 + a^2)$.
 
 Up to lowest order near the horizon, one can then show that
 \begin{equation}
     \chi^2 \equiv \chi^a\chi_a \simeq -\dfrac{\Sigma_+(\theta)(r_+ - r_-)}{(r_+^2 + a^2)^2}(r - r_+),
 \end{equation}
 where $\Sigma_+(\theta) \equiv r_+^2 + a^2\cos^2\theta$. With this expression, the radial vector $\rho^a$ is given in the vicinity of the horizon by 
 \begin{equation}
     \rho^a \simeq 2\kappa (r-r_+)(\partial_r)^a = (\partial_{r^\ast})^a,
 \end{equation}
 which implicitly defines the tortoise coordinate $r_\ast$ via the equation
 \begin{equation}
     2\kappa (r-r_+)\dfrac{\partial r_\ast}{\partial r} = 1 \Rightarrow \kappa r_\ast = \log\sqrt{r-r_+} + C,
 \end{equation}
 where $C$ a $r$-independent integration constant. By picking $C$ as a suitably chosen function of $\theta$, we can set
 
 \begin{equation}
     \kappa x = e^{\kappa r_\ast} = \dfrac{\sqrt{\Sigma_+(\theta)(r_+ - r_-)}}{(r_+^2 + a^2)}(r-r_+)^{1/2},
 \end{equation}
  which makes $\chi^2 \simeq -\kappa^2x^2$ near the horizon, and thus guarantees the Rindler-like terms in the expansion of the metric. The $\theta$-dependence of $\chi^2$ in this case causes the appearance of a term like $x dx\,d\theta$ in the expansion in $(t, x, \theta, \phi)$ coordinates, which can be eliminated by reparametrizing $\theta = \theta' + x^2 F(\theta)$ for a suitably chosen $F(\theta)$. In this way, we have the expansion encoded in \eqref{eqn:Rind} in the $(t, x, \theta', \phi)$ coordinates in Kerr.
  
Note that this construction remains regular in the limit that $a\rightarrow 0$, in which case it gives
the near-horizon coordinates for Schwarzschild.  This differs from HHPS \cite{Haco2018},
whose coordinate system breaks down in the zero spin limit.

\section{Computation of central charge} \label{app:cpp}

The Poisson bracket to evaluate in computing 
the central charge is
\begin{align}
&\;\{H_m, H_n\} 
= \int_{\partial\Sigma}\Big(Q_{\zeta_n}-i_{\zeta_n}\theta[\lie_{\zeta_m} g]\Big) =
\nonumber \\
& \int_{\partial\Sigma} 
\Big( i_{\zeta_m} \theta[\lie_{\zeta_n}g] - 
i_{\zeta_n} \theta[\lie_{\zeta_m}g]
+ i_{\zeta_m}i_{\zeta_n} L - Q_{[\zeta_m,\zeta_n]}\Big), \label{eqn:HHexpanded}
\end{align}
where $L$ is the Einstein-Hilbert Lagrangian.
To arrive at the second line, we use that the generators are field-independent,
$\delta \zeta_n = 0$, that $d Q_\zeta =  \theta[\lie_{\zeta}g] - i_\zeta L$ on shell \cite{Iyer1994a},
and we drop a term $d i_{\zeta_m} Q_{\zeta_n}$ that integrates to zero.  
Then we note from (\ref{eqn:HH}) and (\ref{eqn:KRL}) that 
the central charge just appears as the coefficient of the $m^3$ term,
with all other terms depending linearly on $m,n$.  Examining
(\ref{eqn:HHexpanded}), we find that only the $\theta$ terms 
contain enough derivatives to produce an $m^3$ contribution.  
Evaluation of these terms is aided by noting that 
\begin{equation}
\theta[\lie_{\zeta}g] = \frac{1}{16\pi G} \ep_a\Big[\nabla_b(\nabla^b \zeta^a
-\nabla^a \zeta^b) + 2 R\indices{^a_c}\zeta^c\Big],
\end{equation}
and the Ricci tensor term can be dropped because 
it will not scale as $m^3$. 
The remaining terms involving two derivatives can be evaluated 
in a near horizon expansion, about $w^+ = w^- = 0$.  Note that 
the expression cannot be evaluated directly on the bifurcation surface because
the vector fields $\zeta^a_n$ are singular there for 
$n\neq 0$, as discussed below equation (\ref{eqn:zetxi}). Going through the calculation
using the near-horizon expansion of the metric (\ref{eqn:wpmmetric}),
we arrive at 
\begin{equation} \label{eqn:izIzth}
i_{\zeta_m} \theta[\lie_{\zeta_{-m}} g] =
\frac{-i m(m^2 + \alpha^2)}{8\pi G\alpha(\alpha+\beta)^2}
\Big(\beta+{N_\phi}\Big) \frac{|\psi|}{w^+}\,
dw^+ \wedge \mu ,
\end{equation} 
where $\mu = \sqrt{q}\, d\theta^1\wedge\ldots\wedge d\theta^{(d-3)}$ 
is the volume form for the transverse $\theta^A$ directions. 
The expression for $-i_{\zeta_{-m}} \theta[\lie_{\zeta_m} g]$ is the 
same. The central charge is obtained by integrating the coefficient
of $m^3$ over a cutoff
surface at fixed value of $x$ and $t$,
and then taking the limit to the bifurcation surface, which drops 
all terms regular in $w^+, w^-$. On such a cutoff surface, as 
$\phi$ varies over its range $(0,2\pi)$, $w^+$ goes from 
$w^+_0$ to $w^+_0 e^{2\pi \alpha}$. Since $\displaystyle\int_{w^+_0}^{w^+_0 e^{2\pi\alpha}} \frac{dw^+}{w^+} = 2\pi \alpha$,  we find that 
\begin{equation} 
c_R = \frac{6}{G(\alpha+\beta)^2} \int d\theta^A \sqrt{q}|\psi| 
\Big(\beta + {N_\phi}\Big).
\end{equation}
The first term is simply proportional to the area of the horizon,
$A = 2\pi \int d\theta^A \sqrt q |\psi|$,
while the second term is proportional to $J_H$ (\ref{eqn:JH}).

\section{Wald-Zoupas term} \label{app:WZ}
Here, we provide some details on how to arrive at (\ref{eqn:DcR})
using the Wald-Zoupas prescription identified from the analysis of 
\cite{Chandrasekaran2020}.  The proposed definition
for the quasilocal Hamiltonian on the future horizon $\mathcal{H}^+$ is 
\begin{equation}
\delta H_n = \Omega\big(\delta g_{ab}, \lie_{\zeta_n} g_{ab}\big) 
+\int_{\partial\Sigma} i_{\zeta_n} \mathcal{E},
\end{equation}
where $\partial\Sigma$ lies on a cut of the future horizon,
and the flux $\mathcal{E}$ is a $(d-1)$-form given by
\beq\label{eqn:flux}
\frac{\eta}{16\pi G}\left(2 \varpi^a \chi^b \delta g_{ab} + \delta k -k q^{ab}\delta g_{ab}
+\frac12 k g^{ab}\delta g_{ab}\right) 
\eeq
where $\eta$ is is the volume form induced on $\mathcal{H}^+$ after specifying that 
$\chi_a$ is the null normal, and the inaffinity $k$ and \Hajicek one-form 
$\varpi_a$ are defined below equation (\ref{eqn:cond2}).  Additional terms in the 
flux depending on the expansion and shear vanish since we are working on a Killing 
horizon.
The first term involving the \Hajicek one-form was the Wald-Zoupas counterterm employed
by HHPS in \cite{Haco2018}, but the additional terms are important to ensure 
independence of the choice of auxiliary null vector $n_a$ \cite{Chandrasekaran2020}. 
Note also that this choice of Wald-Zoupas flux coincides with the integrable boundary term
described in section 5.2 of \cite{Chandrasekaran2020}, but differs from the Dirichlet boundary
term considered elsewhere in that work.  
 
The variation $\delta k$ is defined holding $\chi_a$ fixed, which yields the expression
\beq \label{eqn:delzetk}
\delta k = (k n^b -\varpi^b)\chi^a \delta g_{ab} + \chi_c \chi^a n^b\delta \Gamma^c_{ab},
\eeq
and its variation under the transformation generated by $\zeta_n^a$ is 
\beq
\delta_{\zeta_n} k = -\left(w^+\right)^{\frac{in}{\alpha}} \frac{\kappa}{\alpha} n(n-i\alpha) .
\eeq
The \Hajicek form on the future horizon evaluates to
\beq \label{eqn:hajH+}
\varpi_a = N_\phi \nabla_a \phi + N_A \nabla_a \theta^A.
\eeq
The contraction of $\zeta_n^a$ into $\eta$ on $\mathcal{H}^+$ near the bifurcation 
surface is given by
\beq \label{eqn:izeteta}
i_{\zeta_n} \eta = \left(w^+\right)^{\frac{in}{\alpha}} \frac{\alpha}{\kappa(\alpha+\beta)} (\beta -in)\mu,
\eeq
where $\mu$ is the volume form on a cut of constant $x,t$.  

The correction to the central charge is found by computing 
the brackets of the quasilocal Hamiltonians using 
the Barnich-Troessaert bracket \cite{Barnich2011c},
\begin{align}
\{H_m,&\; H_n\} =  \Omega(\lie_{\zeta_m} g, \lie_{\zeta_n} g)  \nonumber \\
& +\int_{\partial\Sigma}\left(
i_{\zeta_n} \mathcal{E}[\lie_{\zeta_m}g] -i_{\zeta_m}\mathcal{E}[\lie_{\zeta_n} g]\right),
\end{align}
where the second line determines the correction to the central charges.  
Evaluating $\{H_m, H_{-m}\}$ 
using the expression (\ref{eqn:flux}) for $\mathcal{E}$ and 
equations (\ref{eqn:delzetk}), (\ref{eqn:hajH+}), and (\ref{eqn:izeteta}) 
then produces the expression (\ref{eqn:DcR}). The analysis on the past horizon 
for the $\xi_n^a$ generators is similar, and yields a correction $\Delta c_L = -\Delta c_R$, 
as is necessary to set the total central charges equal.

\bibliography{bh-soft-refs-v2}

\end{document}